\documentclass[lettersize,journal]{IEEEtran}
\usepackage{amsmath,amsfonts}
\usepackage{algorithm}
\usepackage{array}
\usepackage{textcomp}
\usepackage{stfloats}
\usepackage{url}
\usepackage{verbatim}
\usepackage{graphicx}
\usepackage{cite}
\usepackage{booktabs}
\usepackage{multirow}
\usepackage{makecell}
\usepackage{subfigure}
\usepackage[misc]{ifsym}
\usepackage{listings}
\usepackage{color, xcolor}
\usepackage{algpseudocode}
\usepackage{tcolorbox}

\usepackage{amssymb}
\usepackage{graphicx}
\usepackage{algorithmicx}

\begin{document}

\title{Adaptive Root Cause Localization for Microservice Systems with Multi-Agent Recursion-of-Thought}

\author{
	\IEEEauthorblockN{Lingzhe Zhang, Tong Jia\IEEEauthorrefmark{1}, Kangjin Wang, Weijie Hong, Chiming Duan, \\ Minghua He, and Ying Li\IEEEauthorrefmark{1},~\IEEEmembership{Member,~IEEE}}
	\thanks{Lingzhe Zhang, Tong Jia, Weijie Hong, Chiming Duan, Minghua He, and Ying Li are with Peking University, Beijing, China.}
	\thanks{Email: \{zhang.lingzhe, hongwj, duanchiming, hemh2120\}@stu.pku.edu.cn and \{jia.tong, li.ying\}@pku.edu.cn}
	\thanks{Kangjin Wang is with Alibaba Group, China, e-mail: (kangjin.wkj@alibaba-inc.com)}
	\thanks{* Corresponding author: Tong Jia, e-mail: (jia.tong@pku.edu.cn); Ying Li, e-mail: (li.ying@pku.edu.cn)}
	\thanks{Manuscript received May 28, 2025; revised xxx.}
}

\markboth{Adaptive Root Cause Localization for Microservice Systems with Multi-Agent Recursion-of-Thought}%
{Shell \MakeLowercase{\textit{et al.}}: A Sample Article Using IEEEtran.cls for IEEE Journals}


\maketitle

\begin{abstract}
	As contemporary microservice systems become increasingly popular and complex—often comprising hundreds or even thousands of fine-grained, interdependent subsystems—they are facing more frequent failures. Ensuring system reliability thus demands accurate root cause localization. While traces and metrics have proven to be effective data sources for this task, existing methods either heavily rely on pre-defined schemas, which struggle to adapt to evolving operational contexts, or lack interpretability in their reasoning process, thereby leaving Site Reliability Engineers (SREs) confused. In this paper, we conduct a comprehensive study on how SREs localize the root cause of failures, drawing insights from multiple professional SREs across different organizations. Our investigation reveals that human root cause analysis exhibits three key characteristics: recursiveness, multi-dimensional expansion, and cross-modal reasoning. Motivated by these findings, we introduce RCLAgent, an adaptive root cause localization method for microservice systems that leverages a multi-agent recursion-of-thought framework. RCLAgent employs a novel recursion-of-thought strategy to guide the LLM's reasoning process, effectively integrating data from multiple agents and tool-assisted analysis to accurately pinpoint the root cause. Experimental evaluations on various public datasets demonstrate that RCLAgent achieves superior performance by localizing the root cause using only a single request—outperforming state-of-the-art methods that depend on aggregating multiple requests. These results underscore the effectiveness of RCLAgent in enhancing the efficiency and precision of root cause localization in complex microservice environments.
\end{abstract}


\section{Introduction}

Modern microservice systems have become increasingly complex due to dynamic interactions and evolving runtime environments~\cite{zhou2018fault, zhang2024survey, zhang2025survey}. These systems often consist of hundreds or even thousands of fine-grained, interdependent subsystems, where issues in any one component can easily lead to performance problems at the top level~\cite{mendoncca2019developing, waseem2021design, zhang2024towards, zhang2024time, kang2022separation}. Therefore, to ensure system reliability, it is crucial to localize the root cause of these issues in a timely manner~\cite{zhang2024multivariate, zhang2024reducing}.

However, localizing the root cause of these issues is challenging due to the intricate dependencies between subsystems within microservice systems~\cite{wang2023interdependent, zhang2024failure, yu2024survey, sun2025interpretable, zhu2024hemirca, xie2024microservice, wang2024kgroot, zhang2025log}. Each request typically follows complex invocation chains, with each chain involving multiple components—such as services, service instances, and hosts—and the interactions between them, including service calls, database queries, and other inter-component communications. Moreover, the dynamic nature of these interactions, combined with the heterogeneity of system components, makes pinpointing the root cause a highly non-trivial task.

To enable fine-grained root cause localization of these issues, extensive work leverages both trace and metrics data from software systems. Metrics data provides aggregate insights into system performance (e.g., response times, throughput, and resource utilization), as exemplified by approaches such as Microscope~\cite{lin2018microscope}, which constructs causality graphs and employs a depth-first search strategy to identify front-end anomalies; CIRCA~\cite{li2022causal}, which builds a causal Bayesian network using regression-based hypothesis testing and descendant adjustment to infer problematic components; and RUN~\cite{lin2024root}, which employs time series forecasting for neural Granger causal discovery and integrates a personalized PageRank algorithm to efficiently recommend the top-k root causes. Meanwhile, trace data offers a more granular view by tracking the execution paths of individual requests and inter-component interactions, as demonstrated by MicroRank~\cite{yu2021microrank}, which designs a trace coverage tree to capture dependencies between requests and service instances and applies the PageRank algorithm to score potential root causes; TraceRank~\cite{yu2023tracerank}, which combines spectrum analysis with a PageRank-based random walk to pinpoint abnormal services; CRISP~\cite{zhang2022crisp}, which performs critical path analysis tailored to drill down latency issues; and TraceConstract~\cite{zhang2024trace}, which uses sequence representations along with contrast sequential pattern mining and spectrum analysis to efficiently localize multi-dimensional root causes. Some approaches integrate multimodal data, particularly LLM-based methods. For instance, mABC~\cite{zhang2024mabc} introduces a multi-agent, blockchain-inspired collaboration framework where multiple LLM-based agents follow a structured workflow and coordinate through blockchain-inspired voting mechanisms. RCAgent~\cite{wang2024rcagent} leverages log and code data in a tool-augmented LLM framework to perform root cause analysis tasks such as predicting root causes, identifying solutions, gathering evidence, and determining responsibilities.

Although these root cause localization methods have demonstrated promising outcomes, they still face the following practical challenges when applied to real-world microservice systems:

\begin{itemize}
	\item \textbf{Heavy Reliance on Pre-defined Schemas.} Most approaches depend on pre-defined service causal graphs, fixed statistical models of fault-symptom relationships, or implicitly learned associations from training data—essentially relying on static, one-time analyses. While such schemas provide a baseline capability for root cause localization, they also impose significant limitations on generalizability. The effectiveness of these models is strictly tied to their underlying assumptions; any changes in service dependencies or predefined anomaly models can severely compromise performance. Consequently, these rigid methods struggle to adapt to evolving operational contexts.
	\item \textbf{Lack of Interpretability in the Reasoning Process.} Although some models use deep learning to dynamically capture service dependencies and their correlation with root causes, they often suffer from a lack of transparency compared to graph-based approaches. Given a formatted input, these models can produce a list of potential root causes but typically fail to explain how or why these conclusions are reached. In contrast, an interpretable reasoning process is crucial for Site Reliability Engineers (SREs), as it enables rapid verification of the root cause and facilitates timely remediation.
\end{itemize}

Recognizing this gap, we first conduct an empirical study to investigate how SREs localize the root causes, drawing insights from multiple professional SREs across different organizations. This study reveals three key characteristics of manual root cause analysis: recursiveness (when a deeper-level root cause is identified, SREs iteratively refine their analysis by examining lower-layer manifestations), multi-dimensional expansion (broadening the search across different dimensions, such as pods, services, and infrastructure, to account for all potential root causes), and cross-modal reasoning (once a potential root cause is identified through trace data, it is validated by analyzing fluctuations in relevant metrics).

Building on these insights, we introduce \textbf{RCLAgent}\footnote{Code Repository: \url{https://github.com/LLMLog/RCLAgent}}, an adaptive root cause localization method for microservice systems with multi-agent recursion-of-thought. RCLAgent comprises two types of agents: data agents, which integrate trace and metrics data, and thought agents, which mainly employ recursion-of-thought to iteratively infer and refine the root cause. These agents are coordinated by a central coordinator, which orchestrates the process through three key phases: initial reasoning, critical reflection, and final review.

Our experiments demonstrate that RCLAgent significantly outperforms existing state-of-the-art root cause localization methods. The evaluation results show that RCLAgent delivers superior performance by analyzing a single request, outperforming alternative methods that require multiple requests for analysis. Notably, RCLAgent’s recall@1 even surpasses the recall@10 of competing approaches. Furthermore, when a simple majority voting strategy is employed, RCLAgent’s group ranking MRR exhibits a substantial improvement, surpassing the second-best method by an average of 32.53\%. In summary, the key contributions of this work are as follows:

\begin{itemize}
	\item We conduct a comprehensive study on how SREs localize the root cause an abnormal request. Our findings reveal that human root cause analysis exhibits three key characteristics: recursiveness, multi-dimensional expansion, and cross-modal reasoning.
	\item Inspired by these findings, we propose RCLAgent, an adaptive root cause localization method for microservice systems, leveraging a multi-agent recursion-of-thought framework.
	\item We evaluate RCLAgent on six datasets, demonstrating its effectiveness. Experimental results show that RCLAgent achieves superior performance by analyzing a single request, surpassing state-of-the-art methods that rely on multiple requests for analysis.
\end{itemize}

\section{Background}

In this section, we present the essential background of this paper, including the formal definition of the root cause localization problem, an overview of traditional root cause localization methods, and an introduction to distributed tracing as the primary data source for root cause localization.

\subsection{Root Cause Localization}

Failure diagnosis in distributed systems is generally divided into two categories: failure category classification and root cause localization. The former determines the failure type, such as CPU or memory anomalies, while the latter pinpoints the specific node, service, or pod responsible for the issue. Root cause localization is particularly critical for minimizing downtime and ensuring system stability.

Traditional root cause localization methods typically analyze a collection of anomalous requests within a predefined time window. These methods construct a dependency graph where nodes represent system components, and edges capture causal relationships inferred from request traces or system logs. Graph-based algorithms, such as PageRank, are then applied to rank components based on their likelihood of being the root cause.

\begin{equation}
	C^* = C^* = \arg\max_{C \in \mathcal{C}} s(C, G, M)
	\label{eq: rcl}
\end{equation}

Formally, given a set of anomalous requests $R=\{r_1, r_2, ..., r_n\}$, a component graph $G=(C, E)$ is constructed, where $C$ is the set of components, and $E$ represents their dependencies. The root cause component $C^*$ is identified as Equation~\ref{eq: rcl}, where $s(C, G, M)$ represents the computed score incorporating graph topology and observed anomalies.

\subsection{Distributed Tracing}

To support fine-grained fault diagnosis in software systems, distributed tracing has been widely adopted in industrial environments, becoming an integral part of modern software infrastructures~\cite{wang2022characterizing, yang2022capturing, shen2023network}. A distributed trace provides a detailed execution record of a request as it propagates through the system, capturing timing, dependencies, and performance characteristics. Each trace consists of multiple structured log entries, known as spans, which document individual operations along the request's execution path.

\begin{figure}[htbp]
	\centering
	\includegraphics[width=1\linewidth]{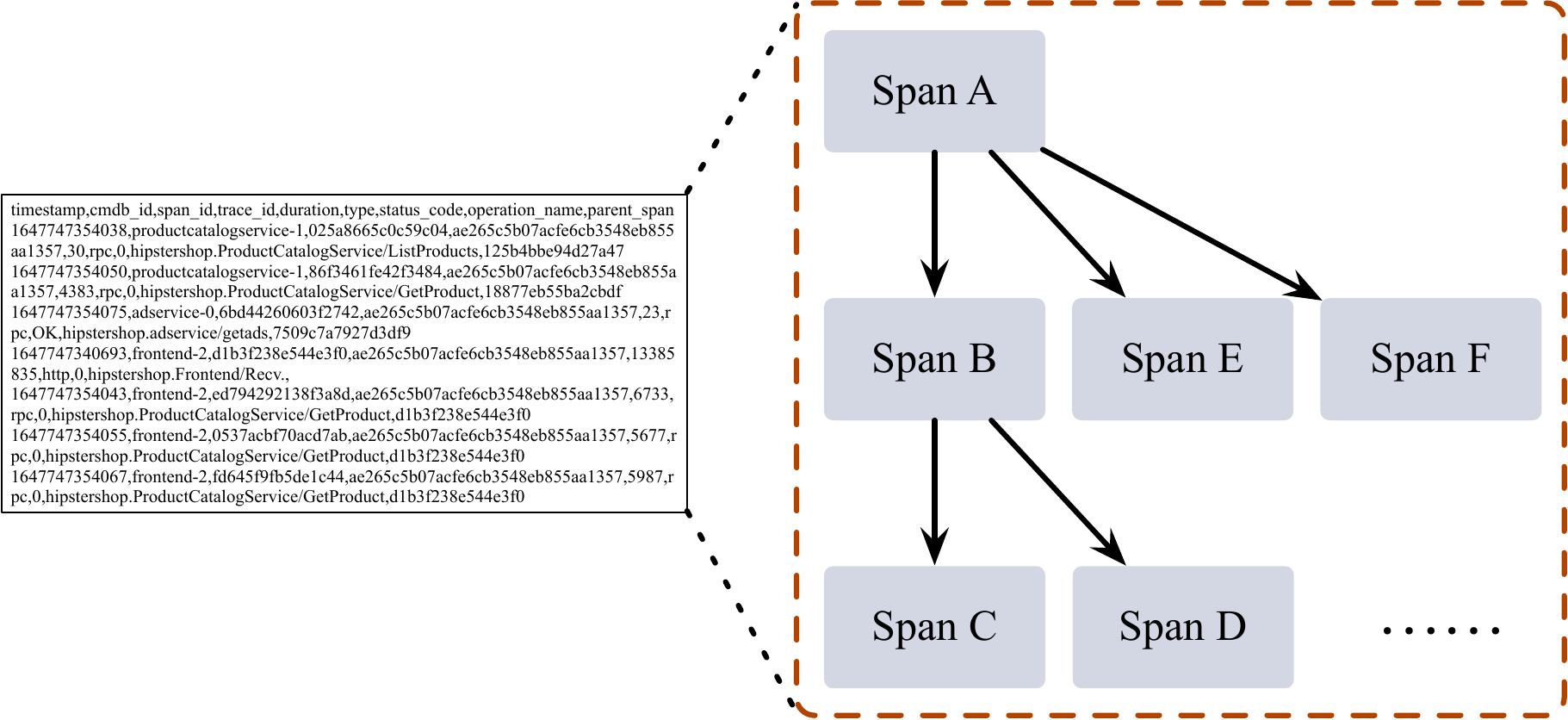}
	\caption{Example of Trace Log}
	\label{fig: tracing}
\end{figure}

As illustrated in Figure~\ref{fig: tracing}, a complete trace represents the end-to-end journey of a request, detailing every intermediate operation and its corresponding latency. Each span records crucial information such as the service name, operations, timestamps, and causal relationships between operations. By analyzing the timing and dependencies of spans, distributed tracing enables precise performance monitoring and facilitates anomaly detection in large-scale distributed systems. This structured representation is particularly valuable for diagnosing latency issues, identifying service bottlenecks, and uncovering failure propagation patterns across microservices.

Each trace contains an entry span that represents the overall execution status of the request. In this paper, we define a request as having an issue if its entry span exhibits an excessively high execution latency—specifically, exceeding 100 times the normal average latency. In such cases, root cause localization is necessary to identify the underlying source of the anomaly.

\section{Empirical Study}

In this section, we conduct a comprehensive study on how SREs localize the root cause of an abnormal request, with insights gathered from multiple professional SREs across different organizations. To systematically analyze this process, we investigate the following three research questions:

\begin{itemize}
	\item \textbf{RQ1:} How do SREs initially narrow down the search space for the root cause?
	\item \textbf{RQ2:} How do they systematically expand the search scope to encompass all potential root causes?
	\item \textbf{RQ3:} How do they differentiate the actual root cause from other potential but non-causal anomalies?
\end{itemize}

\subsection{Recursiveness}

Deep-seated issues often manifest as a request that is exceptionally slow or even fails. To narrow down the search space for the root cause, human analysts begin with the problematic request and examine its trace data to identify the invoked services and executed operations. If anomalies are detected in these services, they further investigate the downstream services they call, recursively refining the search until the true root cause is isolated.

\begin{figure}[htbp]
	\centering
	\includegraphics[width=1\linewidth]{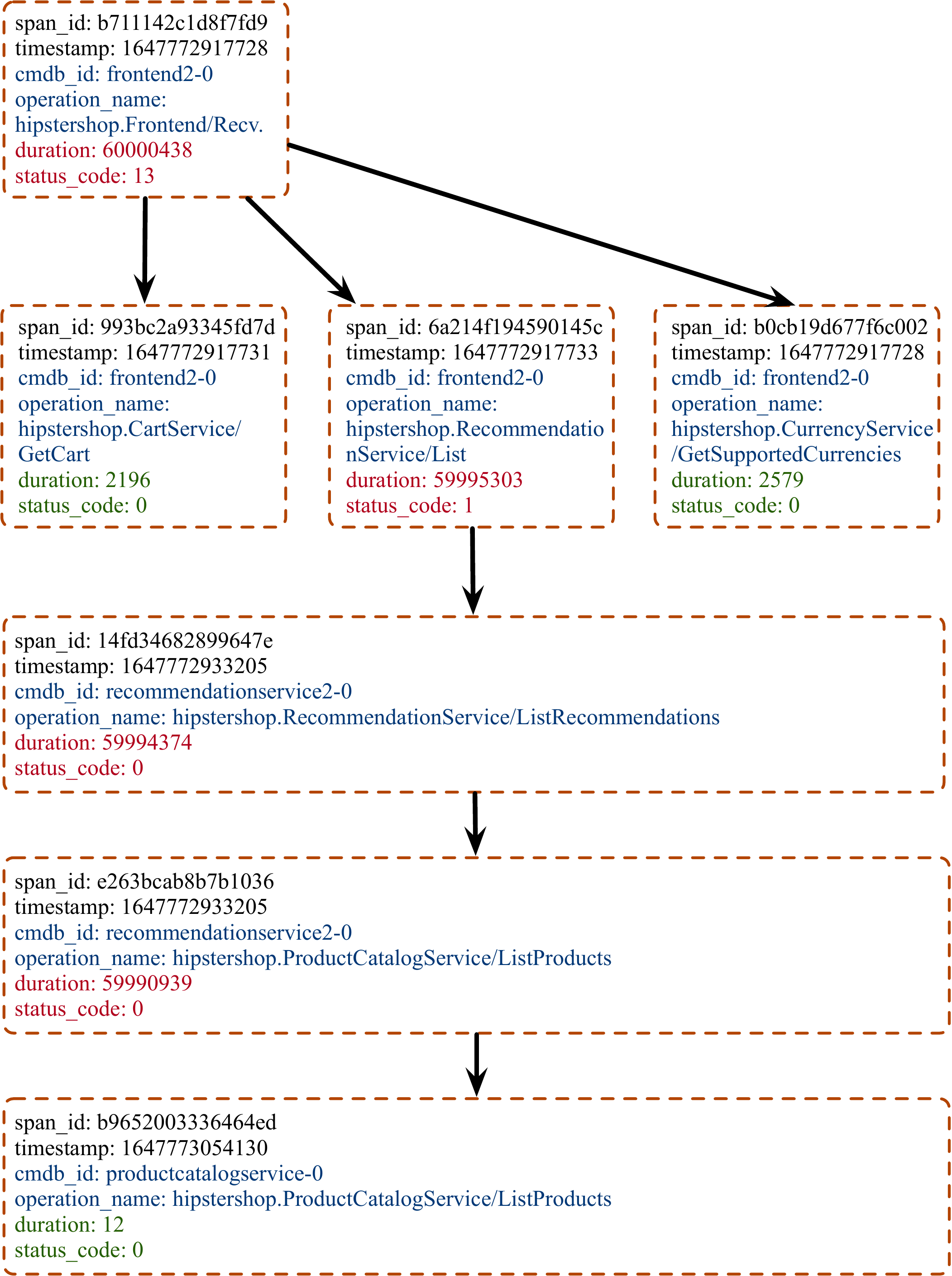}
	\caption{Trace Graph Example-1}
	\label{fig: trace1}
\end{figure}

For instance, we extracted a typical case from the real-world AIOPS 2022 dataset, as shown in Figure~\ref{fig: trace1}. In this case, a request with the cmdb\_id 'frontend2-0' experienced severe timeouts, ultimately resulting in an error (status\_code = 13). As SREs, we first analyzed the trace data for this request and identified three primary operations: CartService/GetCart, RecommendationService/List, and CurrencyService/GetSupportedCurrencies. Notably, only the RecommendationService operation exhibited a timeout, indicating that it was likely responsible for the overall request failure.

We then conducted a deeper investigation into the downstream services and found that the RecommendationService operation with cmdb\_id 'recommendationservice2-0' also experienced a timeout. Further analysis revealed that the subsequent downstream operation, ProductCatalogService, under the same cmdb\_id, also timed out. However, a later operation on the component with cmdb\_id 'productcatalogservice-0' did not exhibit any issues.

This recursive investigation process effectively narrows down the potential root causes to the components with cmdb\_id 'frontend2-0' and 'recommendationservice2-0', significantly reducing the search space for identifying the underlying performance issue.

\begin{center}
	\begin{tcolorbox}[colback=gray!10,
		colframe=black,
		width=\linewidth,
		arc=1mm, auto outer arc,
		boxrule=0.5pt,
		top=2pt, 
		bottom=2pt, 
		left=2pt,
		right=2pt
		]
		\textbf{Summary.} By analyzing trace data, SREs narrow down the search space for the root cause by starting from the entry span of a high-duration request and recursively examining downstream operations until the anomalous component is identified.
	\end{tcolorbox}
\end{center}

\subsection{Multi-Dimensional Expansion}

The previous analysis narrowed the search space by focusing on cmdb\_id, which represents individual pods. However, identifying problematic cmdb\_ids alone is insufficient for pinpointing all root causes. Therefore, SREs expand their analysis beyond individual pods by considering operation\_name, which helps identify the associated services. From the recursive analysis above, it is evident that both recommendationservice and productcatalogservice could also be potential root causes.

\begin{figure}[htbp]
	\centering
	\includegraphics[width=1\linewidth]{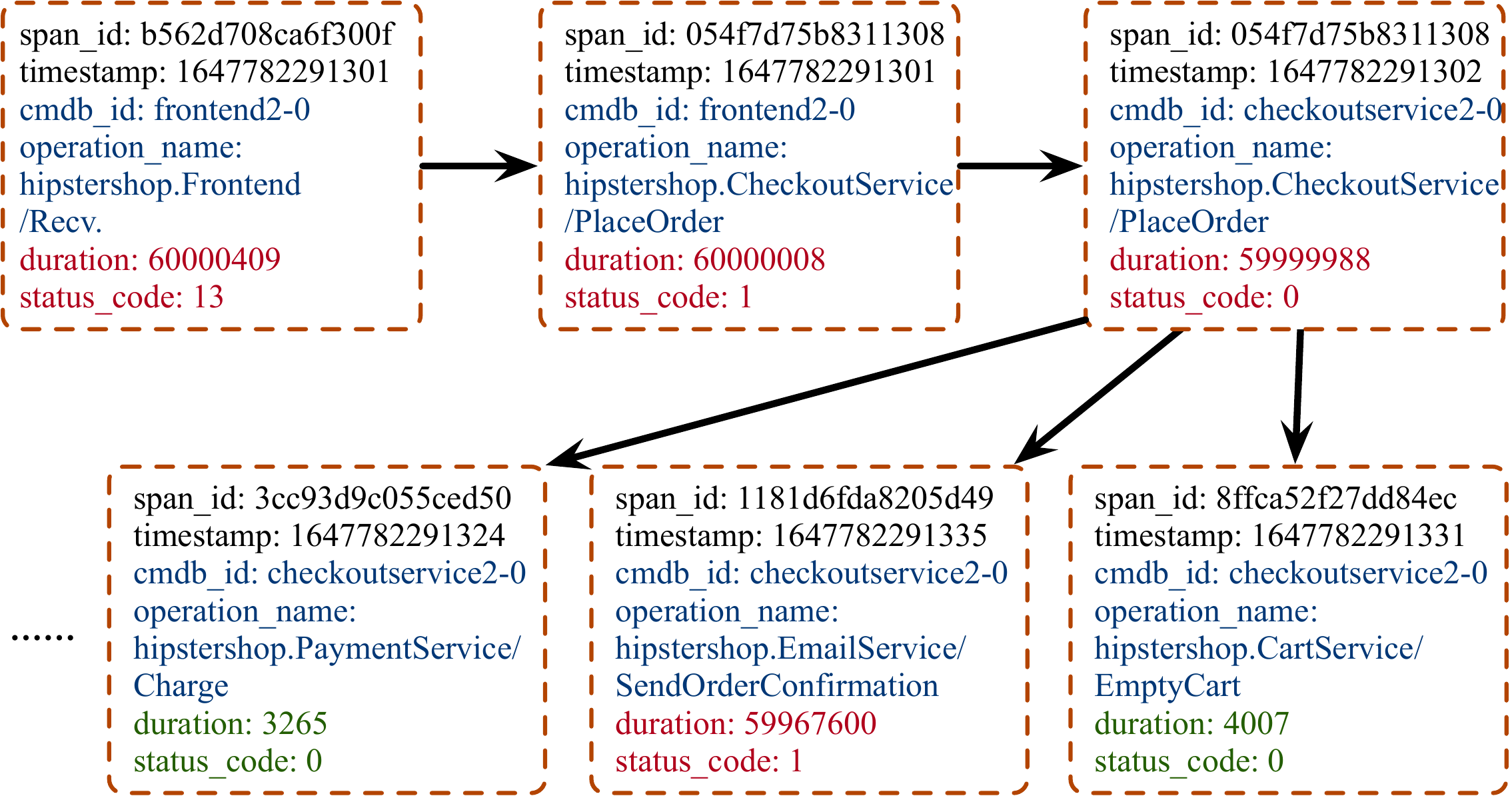}
	\caption{Trace Graph Example-2}
	\label{fig: trace2}
\end{figure}

A more evident example is shown in Figure~\ref{fig: trace2}. In this trace, the root cause is clearly linked to the EmailService operation under cmdb\_id 'checkoutservice2-0'. This suggests that further investigation into emailservice is necessary for a deeper analysis.

Beyond service-level expansion, SREs also consider the underlying physical infrastructure. For instance, in both traces above, 'recommendationservice2-0' and 'checkoutservice2-0' run on node-5, which can be readily obtained in real-time from metrics data. This indicates that further investigation into the physical machine node-5 is required to assess its potential contribution to the issue.

\begin{center}
	\begin{tcolorbox}[colback=gray!10,
		colframe=black,
		width=\linewidth,
		arc=1mm, auto outer arc,
		boxrule=0.5pt,
		top=2pt, 
		bottom=2pt, 
		left=2pt,
		right=2pt
		]
		\textbf{Summary.} SREs systematically expand the scope of root cause analysis across multiple dimensions—correlating anomalous cmdb\_ids with associated services and underlying physical infrastructure—to comprehensively identify all potential sources of issues.
	\end{tcolorbox}
\end{center}

\subsection{Cross-Modal Reasoning}

Through the analysis of recursiveness and multi-dimensional expansion, SREs have identified potential root causes across three dimensions: pod, service, and node. However, the exact root cause remains uncertain. At this stage, SREs often integrate other modalities of data, with metrics data being the most commonly used, to pinpoint the final root cause.

\begin{figure}[htbp]
	\centering
	\includegraphics[width=1\linewidth]{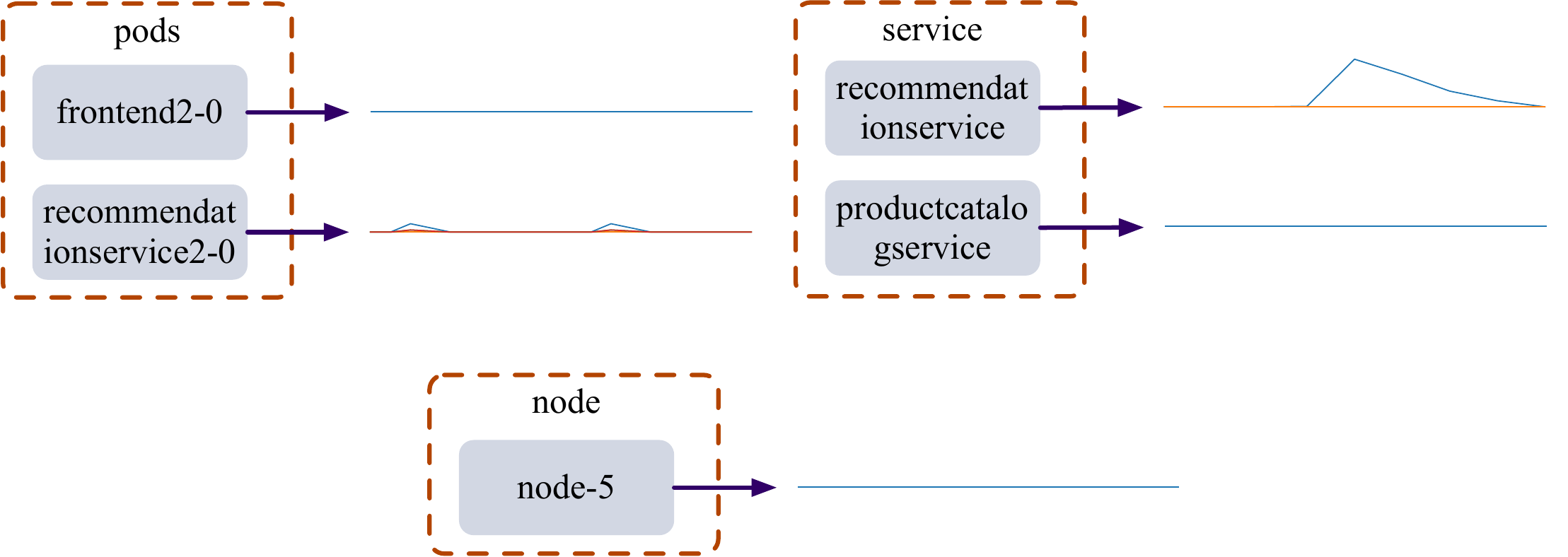}
	\caption{Metrics Inspection of Trace Graph Example-1}
	\label{fig: metrics}
\end{figure}

For example, in Figure~\ref{fig: trace1}, the potentially problematic pod is 'frontend2-0', and 'recommendationservice2-0', the potentially problematic services are 'recommendationservice' and 'productcatalogservice', and the potentially problematic node is 'node-5'. By analyzing the metrics data, as shown in Figure~\ref{fig: metrics}, it becomes clear that there were no fluctuations in 'frontend2-0', 'productcatalogservice', or 'node-5'. However, both 'recommendationservice2-0' and 'recommendationservice' exhibited fluctuations, with 'recommendationservice' showing more significant variations. The fluctuations in 'recommendationservice2-0' were propagated from 'recommendationservice'. Ultimately, the root cause is identified as 'recommendationservice', which aligns with the ground truth—network resource packet corruption in the Kubernetes container.

\begin{center}
	\begin{tcolorbox}[colback=gray!10,
		colframe=black,
		width=\linewidth,
		arc=1mm, auto outer arc,
		boxrule=0.5pt,
		top=2pt, 
		bottom=2pt, 
		left=2pt,
		right=2pt
		]
		\textbf{Summary.} SREs differentiate the actual root cause from other potential anomalies by analyzing other modalities of data, particularly metrics data, and identifying the component with the most significant or earliest fluctuation as the true root cause.
	\end{tcolorbox}
\end{center}

\section{RCLAgent}

Our empirical study demonstrates the process by which human SREs localize the root cause of abnormal requests. However, the time and effort of human experts are valuable, and large language models can effectively simulate this process for automated root cause localization.

\begin{figure*}[htbp]
	\centering
	\includegraphics[width=1\linewidth]{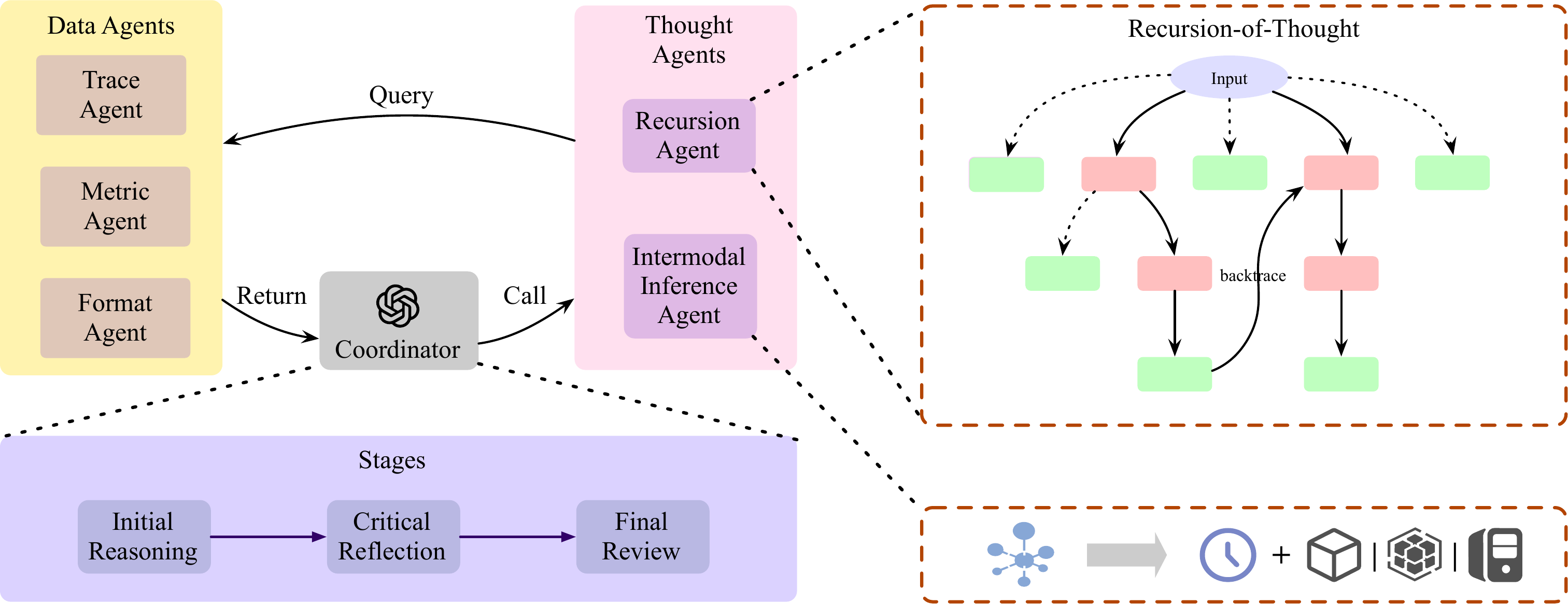}
	\caption{Architecture of RCLAgent}
	\label{fig: architecture}
\end{figure*}

Therefore, in this section, we introduce \textbf{RCLAgent}, an adaptive method for root cause localization for microservice systems through multi-agent recursion-of-thought. Figure~\ref{fig: architecture} illustrates the architecture of RCLAgent.

RCLAgent consists of two types of agents: data agents and thought agents. Data agents are responsible for data retrieval and processing (currently including the Trace Agent, Metric Agent, and Format Agent), while thought agents focus on in-depth reasoning based on the provided data (including the Recursion Agent and Cross-Modal Agent). The coordination between these agents is managed by a central coordinator, which operates through three key phases: initial reasoning, critical reflection, and final review.

\subsection{Data Agents}

Data agents provide various types of data to other agents based on input, along with preprocessing and filtering capabilities. They can encompass any form of data retrieval and processing functions. In this work, we primarily utilize three agents: the Trace Agent, Metrics Agent, and Format Agent. However, in practical applications, the number and types of agents can be flexibly adjusted as needed.

\subsubsection{Trace Agent} 

Trace data is the most fundamental source for root cause localization, as it records the sequence of calls between various services. However, since each request generates a complete invocation path, the volume of trace data can be enormous, making it difficult for large language models to process such extensive context. To address this, the trace agent is designed to filter and retrieve only the relevant subset of trace data.

\begin{equation}
	T(s) = \left\{ \langle t, s', svc, op, d, \sigma \rangle \; \middle| \; s' \in \mathcal{C}(s) \right\}
	\label{eq: trace-agent}
\end{equation}

Formally, given a span identifier $s$, the trace agent returns a set of child spans along with their associated metadata. We denote this function as Equation~\ref{eq: trace-agent}, where $t$ denotes the timestamp, $s'$ represents the child span identifier, $svc$ is the service name, $op$ is the operation name, $d$ is the duration, $\sigma$ is the status code, and $\mathcal{C}(s)$ is the set of all child spans for the given span $s$.

\subsubsection{Metric Agent}

Metrics data represent the runtime state of system components. In a mature microservices environment, there are thousands of distinct metrics continuously recording values, often resulting in a data volume that exceeds that of trace data. Moreover, our empirical study reveals that during system anomalies, most metrics remain stable without significant fluctuations. Consequently, we have developed a specialized Metric Agent that selectively retrieves only those metrics exhibiting notable deviations within a predefined time window for LLM analysis.

\begin{equation}
	|m(t) - \mu_m| > n \times \sigma_m
	\label{eq: n-sigma}
\end{equation}

Formally, given an input timestamp $t_0$ and a target key component $C$ (e.g., a pod or service), let $\mathcal{M}(C)$ denote the set of metrics associated with $C$ and its related components (including the node corresponding to a pod). For each metric $m \in \mathcal{M}(C)$, let $\mu_m$ and $\sigma_m$ represent its historical mean and standard deviation, respectively. The Metric Agent examines the metric values over the interval $[t_0 - \delta, t_0 + \delta]$ and applies an $n$-sigma test: if there exists a time $t$ in this interval, as defined in Equation~\ref{eq: n-sigma}, an anomaly is detected. In such cases, the agent returns the corresponding fluctuation data over a period of length $delta$, as formulated in Equation~\ref{eq: metric-agent}.

\begin{equation}
	Q(t_0, \delta, C) = 
	\left\{ 
	m(t) \;\middle|\; 
	\begin{array}{c}
		m \in \mathcal{M}(C),\; t \in [t_0 - \delta, t_0 + \delta], \\
		|m(t) - \mu_m| > n \times \sigma_m
	\end{array} 
	\right\}
	\label{eq: metric-agent}
\end{equation}

\subsubsection{Format Agent}

Large language models generate extensive natural language outputs that often contain key information. However, this abundance of unstructured text can hinder subsequent automated processing, making it difficult to produce a definitive final result. To address this issue, we designed a Format Agent specifically to summarize the generated text and output it in a standardized format. For instance, when presenting the final results, the Format Agent ensures that exactly two fields are output: one for the root cause (\texttt{root\_cause}) and one for the corresponding explanation (\texttt{reason}).

\subsection{Thought Agents}

Thought agents are designed based on insights from our empirical study, specifically leveraging recursiveness, multi-dimensional expansion, and cross-modal reasoning. Since large language models inherently possess the ability to extract key information from vast amounts of data without excessive deliberation, this section focuses primarily on the design of recursiveness and cross-modal reasoning.

\subsubsection{Intermodal Inference Agent}

Trace data and metrics data exhibit a semantic gap, making direct correlation challenging. To address this, the \textit{Intermodal Inference Agent} is designed to bridge the gap between trace analysis and metrics querying by synthesizing query parameters from multi-modal contextual data. This agent extracts the most relevant keywords (potential root causes) and temporal windows from the trace context. These parameters are then utilized to query the metrics data, ensuring that only those metrics corroborating the anomalies detected in trace data are retrieved.

\begin{equation}
	q = I(X) = (K, \tau)
	\label{eq:intermodal-query}
\end{equation}

Formally, let \( X \) denote the contextual embedding derived from trace analysis. The intermodal inference agent computes a query tuple \( q = (K, \tau) \), as shown in Equation~\ref{eq:intermodal-query}, where \( K \) represents the set of key metric identifiers (i.e., keywords), and \( \tau \) denotes the inferred time window during which anomalies are expected. The function \( I(\cdot) \) encapsulates the agent's inference process, modeling the complex interactions between different data modalities.

\subsubsection{Recursion Agent}

As previously discussed, professional SREs employ a recursive approach when using trace data for root cause localization. To replicate this process, we designed a dedicated Recursion Agent that follows a recursion-of-thought paradigm.

Unlike traditional chain-of-thought reasoning, which follows a linear progression, recursion-of-thought relies on generating step-by-step reasoning instructions based on the information from the previous step. Each generated instruction triggers subordinate agents to retrieve the necessary data for further analysis. Moreover, if a particular analytical path yields no further potential for identifying a root cause, the agent automatically backtraces to a previously encountered candidate that has not been thoroughly inspected, and continues the recursive analysis.

\begin{algorithm}[htbp]
	\caption{Recursion-of-Thought Algorithm}
	\label{alg:recursion-of-thought-trace}
	\begin{algorithmic}[1]
		\Require Trace data \(T\) for a high-duration request.
		\Ensure Formatted potential root cause(s) \(R_f\).
		
		\State Initialize candidate set \(Q \gets \{T_{\text{entry}}\}\)
		\State Initialize potential root cause set \(R \gets \varnothing\).
		
		\While{\(Q \neq \varnothing\)}
		\State Remove a candidate span \(s\) from \(Q\).
		\State Generate reasoning instruction \(I \gets f(s)\)
		\State Retrieve associated data \(D \gets \text{TraceAgent}(I)\).
		
		\If{\(s\) is deemed a potential root cause based on \(D\)}
		\State Compute query tuple \(q \gets I_{\text{intermodal}}(X) = (K, \tau)\)
		\State Retrieve metrics data \(M \gets A_{metric}(q)\).
		
		\If{Metrics analysis confirms anomaly for \(s\)}
		\State \(R \gets R \cup \{s\}\)
		\Else
		\State Discard \(s\)
		\EndIf
		\Else
		\State Let \(C\) be the set of child spans of \(s\).
		\State \(Q \gets Q \cup C\).
		\EndIf
		
		\If{\(s\) yields no further viable path and backtracing is feasible}
		\State Backtrace to a previously uninspected candidate in \(Q\).
		\EndIf
		\EndWhile
		
		\State \(R_f \gets \text{FormatAgent}(R)\)
		\State \Return \(R_f\).
	\end{algorithmic}
\end{algorithm}

In this paper, the recursion-of-thought mechanism is primarily applied to trace data analysis, with metrics data verification used to confirm findings. This process involves the coordinated interaction of all agents in our framework, as illustrated in Algorithm~\ref{alg:recursion-of-thought-trace}.

Specifically, the process begins with the trace agent identifying the entry span of a high-duration request, which serves as the starting point for recursive analysis. At each step, an LLM-based reasoning instruction is generated based on the current candidate span. The trace agent then retrieves the associated trace data, and if the candidate is deemed a potential root cause, the intermodal inference agent synthesizes query parameters (keywords and a temporal window) from the trace context. These parameters are subsequently passed to the metrics agent, which retrieves the corresponding metrics data to verify whether the observed anomaly is corroborated. If the metrics analysis confirms the anomaly, the candidate is added to the potential root cause set; otherwise, the algorithm expands the search by exploring the candidate's child spans. Moreover, if a particular analysis path fails to yield a viable candidate, the system automatically backtraces to previously uninspected candidates for further investigation. Finally, the format agent consolidates and standardizes the identified potential root causes into a structured final output.

\subsection{Coordinator}

Although the recursion agent has outlined the invocation relationships between agents from an algorithmic perspective, the actual process of querying data agents, processing their returned results, and transmitting the information to thought agents for further execution is orchestrated by the coordinator. Moreover, beyond managing these interactions, the coordinator also oversees the execution of Algorithm~\ref{alg:recursion-of-thought-trace} across three distinct phases, ensuring iterative coordination throughout the process.

\subsubsection{Initial Reasoning}

This stage is responsible for initializing the entire inference process. Given an initial task, the model is induced to consider the problem from a recursive perspective and begin a preliminary assessment of potential root causes. To expedite reasoning at this stage, we deliberately hide the presence of the Metrics Agent.

\begin{figure}[htbp]
	\centering
	\begin{tcolorbox}[colback=gray!10, colframe=black, width=\linewidth, arc=1mm, auto outer arc, boxrule=0.5pt, top=2pt, bottom=2pt, left=2pt, right=2pt]
		\textbf{System:} \\
		You are a software operations engineer. Your task is to systematically diagnose and identify the root cause of software failures.\\
		\textbf{User:} \\ 
		Please read the following root trace and identify the corresponding root cause service.\\ 
		A possible approach is to recursively search for the traces you suspect have issues.
		\{$T_{\text{entry}}$\} \\
		You have the following agents to call: \{$A \setminus \{A_{metric}\}$\}
	\end{tcolorbox}
	\caption{The Prompt for Initial Reasoning}
	\label{fig: initial-reasoning}
\end{figure}

As shown in Figure~\ref{fig: initial-reasoning}, the initial prompt, denoted as $P_0$, is constructed from the entry trace $T_{\text{entry}}$ and a set of available agents $A \setminus \{A_{metric}\}$. The initial reasoning state $S_0$ is formally defined as Equation~\ref{eq: initial-reasoning}, where $\Phi(\cdot)$ represents the inference process executed by the large language model.

\begin{equation}
	S_0 = \Phi\Bigl(P_0\bigl(T_{\text{entry}}, \, A \setminus \{A_{metric}\}\bigr)\Bigr)
	\label{eq: initial-reasoning}
\end{equation}

\subsubsection{Critical Reflection}

Initial reasoning can quickly yield a rough judgment of the root cause; however, our experiments reveal that the model tends to terminate its analysis prematurely. For example, as shown in Figure~\ref{fig: trace2}, during the initial reasoning phase the model often identifies the problematic pod as \texttt{checkoutservice2-0} after just two recursive trace queries. In reality, if the model continues the recursion for a few more steps, it discovers that the actual anomaly originates from \texttt{emailservice}. This observation highlights the need for a deeper, more rigorous analysis to avoid early termination and misidentification of the true root cause.

\begin{figure}[htbp]
	\centering
	\begin{tcolorbox}[colback=gray!10,
		colframe=black,
		width=\linewidth,
		arc=1mm, auto outer arc,
		boxrule=0.5pt,
		top=2pt, 
		bottom=2pt, 
		left=2pt,
		right=2pt
		]
		\textbf{Step 1:} \\
		We think it's better to inspect more deeper into the trace tree. So, even you have already define the root cause, you must use the following tool to inspect more deeper to confirm that there are no deeper root cause. \\
		You have the following agents to call: \{$A \setminus \{A_{format}\}$\} \\
		\textbf{Step >1:} \\ 
		Please continue to further identify the root cause service. You may inspect deeper by Trace Agent or combine with Metrics Agent to confirm the root cause.\\
		If you use the Trace Agent, we often find that the root cause originates from a specific downstream trace. If you have already define the root cause service, just call the Format Agent. \\
		You have the following agents to call: \{$A$\}
	\end{tcolorbox}
	\caption{The Prompt for Critical Reflection}
	\label{fig: critical-reflection}
\end{figure}

To address this shortcoming, we introduce a critical reflection stage that forces the model to conduct further investigation before finalizing its decision. In this stage, the model is explicitly prompted to delve deeper into the trace tree, even if an initial root cause candidate has been identified. As depicted in Figure~\ref{fig: critical-reflection}, the model receives a prompt that instructs it to either continue its investigation using the Trace Agent or to incorporate insights from the Metrics Agent for additional verification. This additional layer of scrutiny helps ensure that the model does not overlook deeper anomalies that may be hidden beneath the initial candidate.

\begin{equation}
	S_1 = \mathcal{R}(R_0) = \bigcup_{i=1}^{n} \mathcal{Q}(Q_{i-1}, A)
	\label{eq: critical-reflection}
\end{equation}

Formally, let $R_0$ denote the initial set of potential root causes identified during the initial reasoning stage. The critical reflection stage, denoted as $S_1$, refines $R_0$ by recursively expanding the candidate set. Specifically, we define the refined set $R_f$ as Equation~\ref{eq: critical-reflection}, where $\mathcal{Q}(Q_{i-1}, A)$ represents the expansion of the candidate set $Q_i$ via agent $A$ (such as the Trace or Metrics Agent), and $\mathcal{R}(\cdot)$ encapsulates the recursive refinement process.

\subsubsection{Final Review}

After the critical reflection stage, the RCLAgent performs in-depth inspections of both trace and metrics data. However, in practice, large language models tend to focus more on the recent context when generating results. As a result, potential root causes that appear earlier in the analysis might be overlooked. For example, as shown in Figure~\ref{fig: trace1}, after deeper inspection, the potential root cause could be identified as \texttt{productcatalogservice}. However, the actual root cause in this case lies in the shallower level, \texttt{recommendationservice}, which might be missed if the model prematurely concludes the investigation. This scenario illustrates that, while the initial reasoning might suggest \texttt{recommendationservice}, the critical reflection stage could indicate \texttt{productcatalogservice}, leading to conflicting conclusions.

To address this issue, we introduced a final review stage, where only the Format Agent is invoked. This agent allows the model to review the entire reasoning process, starting from the initial stage, and consolidate its findings to determine the final root cause. By revisiting the prior reasoning, the model ensures that any earlier potential root causes are not overlooked, and the final decision is based on the entire diagnostic process. This structured review helps the model arrive at the most accurate and reliable root cause conclusion.

\begin{figure}[htbp]
	\centering
	\begin{tcolorbox}[colback=gray!10,
		colframe=black,
		width=\linewidth,
		arc=1mm, auto outer arc,
		boxrule=0.5pt,
		top=2pt, 
		bottom=2pt, 
		left=2pt,
		right=2pt
		]
		\textbf{Context:} \\
		Please rethink the above think process and return the correct root cause and reason by the $A_{format}$ agent. \\
		\textbf{Input:} \\ 
		\{Think Process\}
	\end{tcolorbox}
	\caption{The Prompt for Final Review}
	\label{fig: final-review}
\end{figure}

As shown in Figure~\ref{fig: final-review}, the final review stage is implemented as a consolidation process where the reasoning results from both the initial reasoning and critical reflection stages are combined. This is mathematically represented by the Equation~\ref{eq: final-review}, where $\mathcal{R}(R_0)$ represents the initial root cause set from the initial reasoning, and $\mathcal{R}(S_1)$ is the refined set from the critical reflection stage.

\begin{equation}
	R_f = A_{\text{format}}\left( \mathcal{R}(R_0) \cup \mathcal{R}(S_1) \right),
	\label{eq: final-review}
\end{equation}

\section{Evaluation}

To evaluate RCLAgent, we conduct a series of experimental studies to investigate the following research questions:

\begin{itemize} 
	\item \textbf{RQ1:} How accurate is RCLAgent in root cause localization when compared with baseline approaches? 
	\item \textbf{RQ2:} How does the choice of different LLM backbones affect the root cause localization performance of RCLAgent? 
	\item \textbf{RQ3:} What is the contribution of each stage of RCLAgent to its overall accuracy? 
\end{itemize}

\subsection{Experimental Setup}

\subsubsection{Dataset} 

Our experiments are conducted using the AIOPS 2022 companion dataset, which is a large-scale, real-world dataset collected from an actively running, mature microservices-based e-commerce system. This dataset is divided into six distinct subsets. For the sake of clarity and ease of reference in this paper, we label these six subsets as $\mathbf{A}$, $\mathbf{B}$, $\mathbf{\Gamma}$, $\mathbf{\Delta}$, $\mathbf{E}$, and $\mathbf{Z}$.

\subsubsection{Baseline Approaches}

We compared TraceContrast with the following eight root cause localization approaches, which can be categorized into three groups: trace-based, metrics-based, and LLM-based methods.

\textbf{Trace-based:} CRISP~\cite{zhang2022crisp} represents traces as critical paths and applies a lightweight heuristic to identify root cause instances. TraceContrast~\cite{zhang2024trace} utilizes sequence representations, contrastive sequential pattern mining, and spectrum analysis to localize multi-dimensional root causes. TraceRank~\cite{yu2023tracerank} combines spectrum analysis with a PageRank-based random walk algorithm to pinpoint anomalous services. MicroRank~\cite{yu2021microrank} constructs a trace coverage tree to capture dependencies between requests and service instances, leveraging the PageRank algorithm to score potential root causes.

\textbf{Metrics-based:} RUN~\cite{lin2024root} employs time series forecasting for neural Granger causal discovery and integrates a personalized PageRank algorithm to efficiently recommend the top-k root causes. Microscope~\cite{lin2018microscope} builds causality graphs and utilizes a depth-first search strategy to detect front-end anomalies.

\textbf{LLM-based:} mABC~\cite{zhang2024mabc} proposes a multi-agent, blockchain-inspired collaboration framework where multiple LLM-based agents follow a structured workflow and collaborate through blockchain-inspired voting mechanisms. Additionally, we implement a CoT-based Approach~\cite{wei2022chain} which implements a naive chain-of-thought (CoT) reasoning method for root cause localization.

\begin{table*}[tbp]
	\setlength{\tabcolsep}{2.5pt}
	\centering
	\caption{Overall Evaluation Results Compared with non-LLM-Based Methods (Recall@K)}
	\label{tab: overall-recall}
	\begin{tabular}{c|ccc|ccc|ccc|ccc|ccc|ccc|c}
		\toprule
		\multirow{2}{*}{Dataset} & \multicolumn{3}{c|}{CRISP} & \multicolumn{3}{c|}{TraceConstract} & \multicolumn{3}{c|}{TraceRank} & \multicolumn{3}{c|}{MicroRank} & \multicolumn{3}{c|}{RUN} & \multicolumn{3}{c|}{MicroScope} & RCLAgent \\
		& R1 & R5 & R10 & R1 & R5 & R10 & R1 & R5 & R10 & R1 & R5 & R10 & R1 & R5 & R10 & R1 & R5 & R10 & - \\
		\midrule
		$\mathbf{A}$ & 0.00 & 28.14 & 28.14 & 1.69 & 22.71 & 24.07 & 0.68 & 0.68 & 15.93 & 0.00 & 26.10 & 34.93 & 3.73 & 13.22 & 42.03 & 9.68 & 36.92 & \underline{68.46} & \textbf{71.13} \\
		$\mathbf{B}$  & 0.00 & 60.54 & 61.88 & 56.05 & 75.78 & 75.78 & 75.78 & 75.78 & 75.78 & 0.90 & 60.54 & \underline{76.23} & 0.00 & 0.00 & 0.00 & 1.47 & 7.35 & 14.22 & \textbf{78.57} \\
		$\mathbf{\Gamma}$ & 2.51 & 48.62 & 51.62 & 46.35 & 78.92 & 88.98 & 23.95 & 43.35 & 56.65 & 15.45 & 80.84 & \underline{89.70} & 23.83 & 24.55 & 24.55 & 20.95 & 59.76 & 81.18 & \textbf{90.24} \\
		$\mathbf{\Delta}$ & 1.15 & 52.87 & 56.05 & 2.19 & 2.92 & 3.13 & 59.44 & 60.06 & \underline{63.48} & 2.82 & 3.34 & 3.34 & 0.42 & 14.18 & 17.52 & 7.61 & 20.75 & 26.07 & \textbf{64.34} \\
		$\mathbf{E}$ & 15.20 & 62.33 & \underline{63.06} & 33.26 & 34.14 & 35.68 & 23.79 & 51.76 & 59.69 & 27.09 & 35.68 & 35.68 & 0.00 & 6.39 & 31.28 & 8.81 & 37.44 & 54.19 & \textbf{65.42} \\
		$\mathbf{Z}$ & 4.68 & 40.76 & 45.32 & 4.81 & 11.65 & 17.22 & 34.94 & 35.70 & 39.75 & 4.43 & 16.96 & 17.47 & 0.13 & 12.91 & 39.49 & 6.86 & 37.98 & \underline{63.39} & \textbf{72.41} \\
		\bottomrule
	\end{tabular}
\end{table*}

\subsubsection{Evaluation Metrics}

We use the top-k recall (Recall@k) and mean reciprocal rank (MRR) to evaluate the accuracy of root cause localization following existing works~\cite{zhang2024trace, yu2021microrank, lin2018microscope, zhang2024survey}.

\begin{itemize}
	\item \textbf{Recall@k:} Measures the likelihood that the true root cause appears within the top-k results in the ranked list. Specifically, it indicates whether the root cause is found within the first $k$ predictions. In this paper, we evaluate Recall@1, Recall@5, and Recall@10.
	\item \textbf{MRR:} is the multiplicative inverse of the rank of the root cause in the result list. If the root cause is not included in the top-10 result list, the rank can be regarded as positive infinity. Given a set of fault instances $A$, $Rank_i$ is the $i$ rank of the root cause in the returned list of the $i$th fault instance, MRR is calculated by Equation~\ref{eq: mrr}.
\end{itemize}

\begin{equation}
	MRR = \frac{1}{|A|} \sum_{i=1}^{|A|} \frac{1}{Rank_i}
	\label{eq: mrr}
\end{equation}

Since many existing methods are limited to localizing the root cause only at the pod level, we assume that if these methods correctly identify the pod, the corresponding service to which the pod belongs is considered the root cause, and we treat such predictions as correct.

\subsubsection{Implementation and Settings}

We implement RCLAgent in Python 3.10. Unless otherwise specified, we use Claude-3.5 Sonnet as our LLM engine and set the n-sigma threshold in the metrics agent to $n = 3$. All experiments are conducted on a Linux server equipped with 24 Intel(R) Xeon(R) processors (2.90GHz), 400GB RAM, and two A800 GPUs with 80GB of GPU memory, running CentOS 8.

\subsection{Overall Evaluation}

We first compare RCLAgent with trace-based methods and metrics-based methods. As shown in Table~\ref{tab: overall-recall}, all of these SOTA methods rely on initial service relationships and predefined causal graph construction approaches. Therefore, they require collecting multiple requests and constructing a causal graph before ranking potential root causes based on their weights. In contrast, RCLAgent only analyzes a single request and directly identifies the exact root cause. Thus, the results presented for RCLAgent in the table correspond to Recall@1 (R1), since it determines the exact root cause with just one request.

From the table, we observe that even when analyzing only a single request to identify the exact root cause, RCLAgent outperforms the SOTA methods' Recall@10 (R10) results by approximately 2\% to 9\%. Furthermore, compared to other methods achieving the same Recall@1 (R1) performance, RCLAgent shows an average improvement of 30.44\%. This demonstrates the superior effectiveness of RCLAgent, even when working with a single request.

\begin{table}[htb]
	\setlength{\tabcolsep}{5pt}
	\centering
	\caption{Evaluation Results Compared with LLM-Based Methods}
	\label{tab: llm-comparison}
	\begin{tabular}{c|cccccc}
		\toprule
		Approach & $\mathbf{A}$ & $\mathbf{B}$ & $\mathbf{\Gamma}$ & $\mathbf{\Delta}$ & $\mathbf{E}$ & $\mathbf{Z}$ \\
		\midrule
		CoT & 11.90 & 21.43 & 39.95 & 16.25 & 20.20 & 25.47 \\
		mABC & 51.87 & 62.22 & 74.23 & 56.35 & 52.08 & 51.65 \\
		RCLAgent & \textbf{71.13} & \textbf{78.57} & \textbf{90.24} & \textbf{64.24} & \textbf{65.42} & \textbf{72.41} \\
		\bottomrule
	\end{tabular}
\end{table}

Next, we compare the performance of RCLAgent with LLM-based methods. As shown in Table~\ref{tab: llm-comparison}, when evaluating on Recall@1 (R1), RCLAgent outperforms mABC by an average of 15.60\% in accuracy. Additionally, mABC combines log data analysis alongside trace and metric data, and with a higher number of agents involved, its inference speed is slower compared to RCLAgent. In comparison with the naive CoT strategy, RCLAgent demonstrates a much more significant advantage, outperforming it by up to 60\% on certain datasets. This result highlights that root cause localization is not a problem that can be easily solved by merely using LLMs, further emphasizing the necessity of the design of RCLAgent.

\subsection{Group Ranking Evaluation}

The previous experiments demonstrated that RCLAgent, by analyzing a single request to provide the exact root cause, can outperform SOTA methods. To further assess the superiority of RCLAgent, we introduced a naive majority voting algorithm. Specifically, for a set of multiple requests within a time window, RCLAgent analyzes each request to identify its corresponding root cause. The root causes are then aggregated using a majority voting mechanism to determine the final root cause, in a manner analogous to SOTA methods, which rely on constructing a causal graph and ranking potential root causes.

Formally, the majority voting procedure can be described as follows. Let $\{r_1, r_2, ..., r_k\}$ represent the root causes identified for $k$ requests within a time window. The majority vote is given by Equation~\ref{eq: majority-voting}, where $\hat{R}$ is the final predicted root cause, $\mathbb{I}(r_i = r)$ is the indicator function that equals 1 if the $i$-th root cause $r_i$ matches the candidate root cause $r$, and 0 otherwise. The root cause $\hat{R}$ is the one that receives the highest number of votes.

\begin{equation}
	\hat{R} = \text{argmax}_r \left( \sum_{i=1}^{k} \mathbb{I}(r_i = r) \right)
	\label{eq: majority-voting}
\end{equation}

\begin{table}[htb]
	\setlength{\tabcolsep}{4pt}
	\centering
	\caption{MRR Comparison}
	\label{tab: mrr-comparison}
	\begin{tabular}{c|cccccc}
		\toprule
		Approach & $\mathbf{A}$ & $\mathbf{B}$ & $\mathbf{\Gamma}$ & $\mathbf{\Delta}$ & $\mathbf{E}$ & $\mathbf{Z}$ \\
		\midrule
		CRISP & 8.27 & 20.13 & 18.13 & 17.34 & 31.08 & 17.14 \\
		TraceConstract & 13.07 & 65.74 & \underline{58.55} & 2.48 & 33.77 & 8.15 \\
		TraceRank & 6.26 & \underline{76.76} & 34.41 & \underline{61.54} & \underline{35.79} & \underline{38.36} \\
		MicroRank & 11.38 & 18.12 & 38.10 & 2.98 & 30.81 & 9.15 \\
		RUN & 11.72 & 3.12 & 25.65 & 5.62 & 7.58 & 8.95 \\
		MicroScope & \underline{23.76} & 4.55 & 37.46 & 13.24 & 21.38 & 21.33 \\
		RCLAgent & \textbf{77.63} & \textbf{78.57} & \textbf{94.44} & \textbf{84.98} & \textbf{76.02} & \textbf{78.31} \\
		\bottomrule
	\end{tabular}
\end{table}

After applying the aforementioned majority voting algorithm, we conducted a detailed analysis of RCLAgent and the SOTA methods in terms of MRR, as shown in Table~\ref{tab: mrr-comparison}. From the table, it is evident that RCLAgent outperforms all the SOTA methods in terms of MRR. On average, RCLAgent achieves a 32.53\% improvement over the second-best method. This demonstrates the superiority of RCLAgent in the group ranking evaluation. Its excellent performance in Recall@1 (R1) ensures that even with a simple majority voting algorithm, RCLAgent delivers impressive results.

\subsection{LLM Backbone Impact}

Next, we address the second research question. To evaluate the performance of RCLAgent across different LLM backbones, we conducted experiments with four additional state-of-the-art LLMs beyond Claude-3.5-Sonnet: DeepSeek-R1-Qwen, which is the Qwen2.5-32B model fine-tuned on distilled data released by DeepSeek; Qwen-2.5-Max and Qwen-2.5-Plus, both of which are closed-source models accessed via API; and Llama-3.1-70B, the original open-source version released by Meta.

\begin{table}[htb]
	\setlength{\tabcolsep}{3pt}
	\centering
	\caption{RCLAgent with different LLMs}
	\label{tab: llm-impact}
	\begin{tabular}{c|cccccc}
		\toprule
		Model & $\mathbf{A}$ & $\mathbf{B}$ & $\mathbf{\Gamma}$ & $\mathbf{\Delta}$ & $\mathbf{E}$ & $\mathbf{Z}$ \\
		\midrule
		Claude-3.5-sonnet & \textbf{71.13} & \textbf{78.57} & \textbf{90.24} & \textbf{64.34} & \textbf{65.42} & \textbf{72.41} \\
		DeepSeek-R1-qwen & 33.33 & 58.73 & \underline{67.55} & \underline{69.17} & \underline{45.06} & \underline{61.36} \\
		Qwen-2.5-max & 13.17 & 21.23 & 33.66 & 13.17 & 33.66 & 13.68 \\
		Qwen-2.5-plus & \underline{39.79} & \underline{70.45} & 52.78 & 49.48 & 35.46 & 52.06 \\
		Llama-3.1-70B & 7.92 & 22.42 & 17.51 & 18.14 & 6.97 & 8.76 \\
		\bottomrule
	\end{tabular}
\end{table}

As shown in Table~\ref{tab: llm-impact}, the performance of RCLAgent is closely tied to the reasoning capabilities of the underlying LLM backbone. In our experiments, Claude-3.5-Sonnet consistently achieved the best results, followed by DeepSeek-R1-Qwen. However, due to its relatively smaller size (32B), DeepSeek-R1-Qwen underperformed on datasets A and B compared to Qwen-2.5-Plus. Overall, RCLAgent with Claude-3.5-Sonnet outperformed the second-best model by an average of 14.79\%.

Additionally, we experimented with other models such as GPT-4o and Llama-3-70B, but their performance was even lower than the weakest models listed in Table~\ref{tab: llm-impact}. These findings highlight that RCLAgent effectively leverages the semantic understanding and logical reasoning capabilities of large language models, achieving better results when paired with more powerful inference engines.

\subsection{Ablation Study}

Then, to address the third research question, we recorded the complete reasoning path of RCLAgent during the experiments. Additionally, we extracted intermediate results at three distinct stages and evaluated their effectiveness in root cause localization.

\begin{figure}[htbp]
	\centering
	\includegraphics[width=1\linewidth]{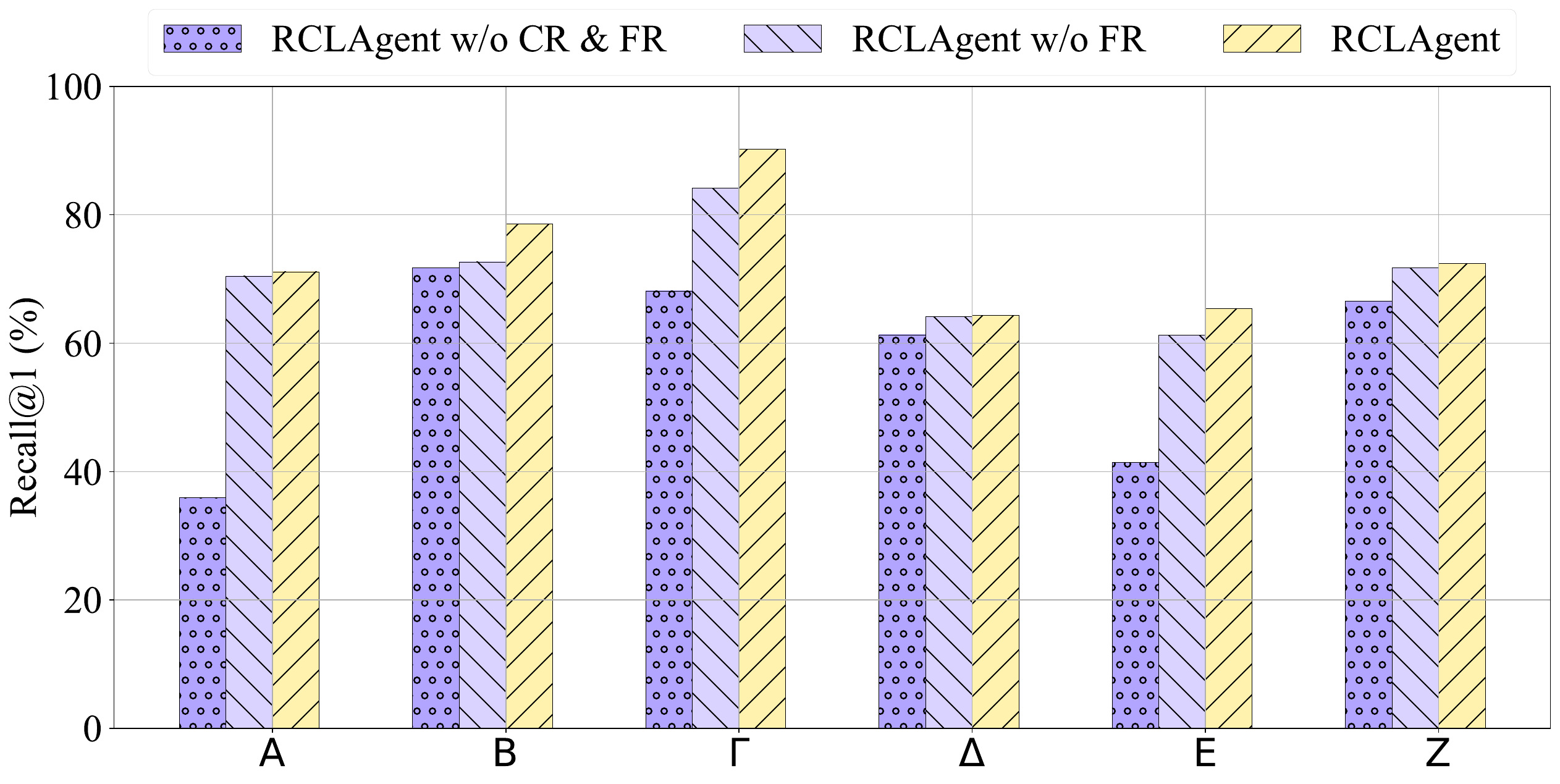}
	\caption{Ablation Experiment}
	\label{fig: ablation}
\end{figure}

As shown in Figure~\ref{fig: ablation}, CR denotes Critical Reflection, and FR represents Final Review. The figure demonstrates that these three stages significantly enhance RCLAgent's root cause localization performance. For instance, on dataset $\mathbf{A}$, applying Critical Reflection improved the results of Initial Reasoning by approximately 34.51\%. Similarly, Final Review consistently contributed to performance gains, albeit to a lesser extent than Critical Reflection. On datasets $\mathbf{B}$ and $\mathbf{\Gamma}$, Final Review further improved the results by approximately 6\% beyond the gains from Critical Reflection.

Overall, Initial Reasoning alone achieved an R1 (Recall@1) score of approximately 57.52\%. Incorporating Critical Reflection led to an average improvement of 13.31\%, while Final Review provided an additional 2.96\% boost. These findings highlight the effectiveness and rationality of RCLAgent’s three-stage design.

\section{Threats to Validity}

Despite the promising results demonstrated by RCLAgent, several limitations should be acknowledged. First, the implementation and configuration of baseline approaches: CRISP, MicroRank, and mABC have publicly available source code, which we directly utilized. For other state-of-the-art methods, we implemented them based on the descriptions in their respective papers. After implementation, we fine-tuned the results and selected the best configurations for all baselines through experimentation. Second, our evaluation primarily relies on the AIOps 2022 dataset, which, while extensive and derived from real-world microservice systems, may not fully capture the diversity of failure scenarios encountered in different operational contexts. Third, although our empirical study benefits from insights provided by multiple professional SREs, the inherent subjectivity in their diagnostic processes may introduce bias.

\section{Related Work}

\subsection{Trace-Based Root Cause Localization}

With the advancement of distributed tracing and its supporting infrastructure, researchers have explored various trace-based techniques for root cause localization. Several studies~\cite{li2022enjoy, luo2021characterizing, zhou2018fault} have analyzed the role of distributed tracing in diagnosing failures in large-scale microservices systems, highlighting the importance of automated trace analysis for effective root cause localization.

Recently, machine learning techniques have been widely adopted for trace-based root cause localization. For instance, Zhou et al.~\cite{zhou2019latent} proposed MEPFL, a supervised learning-based approach. Gan et al.~\cite{gan2019seer} introduced Seer, a model leveraging CNNs and LSTMs for root cause analysis. Liu et al.~\cite{liu2020unsupervised} proposed TraceAnomaly, an unsupervised approach for anomaly detection in traces. Additionally, Gan et al.~\cite{gan2021sage} developed Sage, which utilizes graph neural networks for root cause localization.

Beyond machine learning-based methods, some researchers have extended spectrum analysis techniques to improve the practicality of trace-based root cause localization. For example, Yu et al.~\cite{yu2021microrank} proposed MicroRank, which was later extended into TraceRank~\cite{yu2023tracerank} by integrating spectrum analysis with a random walk-based method. Li et al.~\cite{li2021practical} combined spectrum analysis with frequent pattern mining to identify root cause service instances. More recently, Zhang et al.~\cite{zhang2024trace} introduced TraceConstruct, which leverages sequence representations along with contrast sequential pattern mining and spectrum analysis to efficiently localize multi-dimensional root causes.

\subsection{LLM-based Failure Management}

Large language models, with their advanced semantic understanding and logical reasoning capabilities, have significantly improved the field of failure management~\cite{zhang2024survey} and are increasingly becoming a focal point of research. Numerous LLM-based approaches have been proposed to address various aspects of failure management, including anomaly detection, failure diagnosis, and automated mitigation~\cite{rasul2023lag, liu2024timer, das2024decoder, shi2023shellgpt, liu2024anomalyllm, liu2024unitime, guo2023owl, liu2024loglm, chen2024automatic, jiang2024xpert, zhang2024lm, hamadanian2023holistic, pan2024raglog, zhang2024lograg, zhang2025xraglog, zhang2025scalalog, zhang2025agentfm, zhang2025thinkfl, zhang2025logdb, eagerlog, midlog, famos, logcae, llmelog, afalog, aclog, zhang2025surveyparallel}.

Some studies have developed foundation models specifically for failure management. For example, Lag-Llama~\cite{rasul2023lag}, Timer~\cite{liu2024timer}, and TimesFM~\cite{das2024decoder} pretrain foundation models for metrics-based anomaly detection. Similarly, ShellGPT~\cite{shi2023shellgpt} trains a model capable of automatically generating shell scripts for automated mitigation.

Other approaches adopt fine-tuning strategies to tailor LLMs for failure management tasks. For instance, AnomalyLLM~\cite{liu2024anomalyllm} and UniTime~\cite{liu2024unitime} employ full fine-tuning for anomaly detection, while OWL~\cite{guo2023owl} and LogLM~\cite{liu2024loglm} leverage parameter-efficient fine-tuning techniques to build log analysis models.

Since these fine-tuning approaches require significant computational resources and time, an increasing number of methods, including our work, rely on prompt-based techniques. For example, RCACopilot~\cite{chen2024automatic} and Xpert~\cite{jiang2024xpert} utilize in-context learning (ICL) to structure diagnostic processes, ensuring accurate root cause analysis. LM-PACE~\cite{zhang2024lm} applies chain-of-thought (CoT) reasoning to enhance GPT-4’s ability to analyze incident reports, while Hamadanian et al.~\cite{hamadanian2023holistic} extend this approach to generate mitigation solutions from incident reports. Additionally, RAGLog~\cite{pan2024raglog} and LogRAG~\cite{zhang2024lograg} use retrieval-augmented generation (RAG) to enhance log-based anomaly detection through historical log retrieval.

\section{Conclusion}

In this paper, we study the problem of adaptive root cause localization for microservice systems. Initially, we conduct a comprehensive study on how SREs localize the root causes of an abnormal request. This study reveals that human root cause analysis exhibits three key characteristics: recursiveness, multi-dimensional expansion, and cross-modal reasoning. Based on our study, we introduce RCLAgent, an adaptive root cause localization method for microservice systems that leverages a multi-agent recursion-of-thought framework. RCLAgent employs a novel recursion-of-thought approach to guide the LLM's reasoning process, effectively utilizing data from multiple agents and tool-assisted analysis to achieve accurate root cause localization. Our experimental evaluations on various public datasets demonstrate that RCLAgent achieves superior performance by analyzing a single request, surpassing state-of-the-art methods that rely on multiple requests for analysis.

In future work, we will further explore how to achieve more accurate and efficient root cause localization using smaller-scale models. Additionally, we are considering extending our approach to cover the entire failure management process.

\bibliographystyle{IEEEtran}
\bibliography{mylib}

\end{document}